\documentclass[12pt]{iopart}

\usepackage{hyperref}
\usepackage{graphicx}
\usepackage{rotating}
\usepackage{array}

\def\be{\begin{equation}}
\def\ee{\end{equation}}
\def\bea{\begin{eqnarray}}
\def\eea{\end{eqnarray}}
\def\ahalf{{\textstyle{1\over2}}}

\def\ie{{\it i.e.\,}}
\def\etal{{\it et al.   }}

\def\<{\langle}
\def\>{\rangle}

\begin{document}

\title{Unusual Bound States in Quantum Chains}
\author{E. Sadurn\'i $^{1}$}

\address{$^1$Instituto de F\'isica, Benem\'erita Universidad Aut\'onoma de Puebla,
Apartado Postal J-48, 72570 Puebla, M\'exico}

\eads{ \mailto{sadurni@ifuap.buap.mx}}

\begin{abstract}

The existence of bound states in quantum mechanics with no classical counterpart has been a subject of interest for a long time. Cross-wires and cavities connected to infinite leads are typical examples in which open geometries with bulges support bound solutions, in two or more dimensions. Here we find that the role of topology can be even more important than space availability, by showing the existence of bound solutions in one-dimensional systems such as quantum cross-chains and, in general, chains tied in geometries without loops. It is shown that these examples of unusual binding can be solved analitically for energies and eigenvectors. An experimental proposal is given in the form of tight-binding arrays of electromagnetic resonators, as the effects in question are of a wave-like nature.

\end{abstract}


\maketitle

\pagestyle{plain}

\tableofcontents

\pagestyle{headings}

\section{Introduction}

\begin{figure}[!h] \begin{center} \begin{tabular}{cc} \includegraphics[scale=0.35]{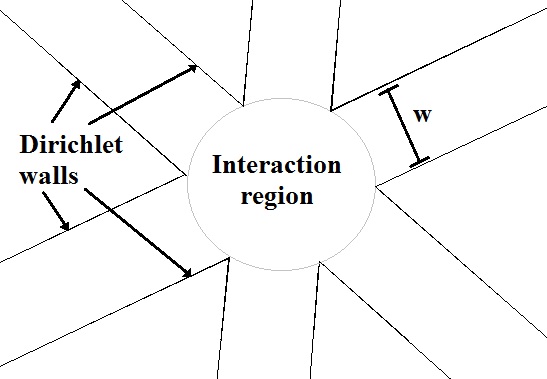} &  \includegraphics[scale=0.35]{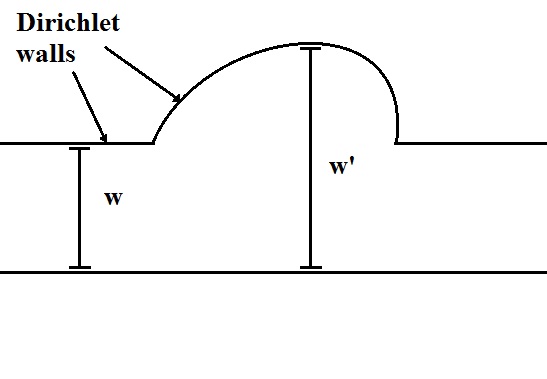} \end{tabular} \end{center} 

\caption{ \label{fig:1} Boundary value problems for the stationary Schr\"odinger equation in open geometries. These two examples support bound states due to the presence of bulges and cavities. The six-wire configuration (left panel) supports bound states in the circular shape at the centre and whose energies lie below the propagating threshold, defined by the lowest energy mode in the straight leads $E_t=\pi^2/w^2$. In the case of the wire with a bulge (right panel), the increase in thickness from $w$ to $w'$ produces the same effect.} \end{figure}

Trapping mechanisms of particles are ubiquitous in physical applications and usually entail the presence of external potentials which rise in all possible directions. There are, however, a few examples in which this is not the case. Several years ago, Schult, Ravenhall and Wyld \cite{1} found numerical evidence for the existence of bound states in cross-wires, despite of the fact that the corresponding boundary-value problem is defined by an open geometry. In such a situation, classical particles always have the chance to escape. These unusual bound states are a purely wave-like phenomenon and they are but an example of the many interesting trapping effects that wave dynamics provides (e.g. bound states dissolved in the continuum \cite{2, 3, 4}, bound solutions in smoothly curved wave guides \cite{5, 6, 7, 8}, etc). Over the years, it has been recognized \cite{9, 10, 11} that bound states in open geometries are supported by an increase of available space, either due to a large area in a cavity connected to infinite leads or due to the presence of bulges in tubes \cite{12}). The mechanism can be understood as the contribution of long wavelengths inside a large region corresponding to evanescent modes along the tubes connected to the system. There is a clear separation between the evanescent states of low energy (typically forming a discrete spectrum) and the solutions above the propagating threshold, defined by the largest wavelength that is able to escape to infinity through the guides (the continuum of the system). See figure \ref{fig:1}.

This sort of intuition, of course, is not a recent development. Some scattering models of nuclear physics depict a stable composite as a bound state in some interaction region, while the possible outcomes of a reaction (decay) correspond to the propagating modes in the leads \cite{13, 14}. In this context, it is well-known that an appropriate way to deal with such problems is through the study of the analytic properties of the scattering matrix or the reaction matrix as a function of energy and that the bound state energies can be identified as a specific type of poles \cite{15, 16}.

 \begin{figure}[!h] \begin{center} \begin{tabular}{cc}  \includegraphics[scale=0.35]{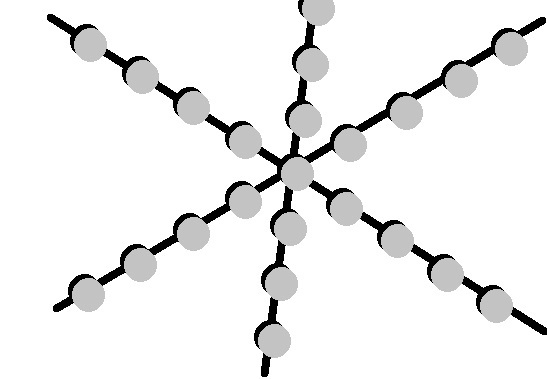} & \includegraphics[scale=0.35]{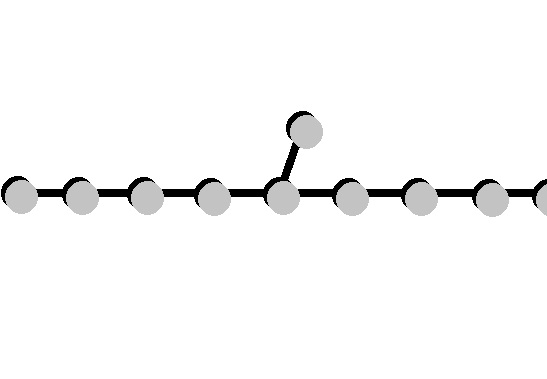}  \end{tabular} \end{center} 

\caption{ \label{fig:2} Tight-binding chains as caricatures of the previous examples with wires in figure \ref{fig:1}. These configurations support also bound states, despite of the fact that there is only one site at the centre of the six-chain configuration (left panel) and only one site stemming out of the linear chain (right panel). In the latter case, the effect can be thought as the binding emerging around defects in periodic systems.} \end{figure}

In this paper, we present yet another type of bound states which can be viewed as a certain limit of the aforementioned wires, but which seems to contradict the intuitive arguments related to an increase of available space for the waves inside. When our system with boundaries becomes one-dimensional and discrete, we must abandon the ideas given above and enter into the realm of chains, the tight-binding approximation being an important example. Such chains, as we shall see, support a fixed number of bound states (two states) when they are semi-infinite and they are coupled to each other by their ends in the form of stars  (henceforth referred to as the {\it arms\ }of the array). In particular, they can be arranged in the form of a cross, just as its continuous counterpart studied in \cite{1}. See figure \ref{fig:2}.

Interestingly, such crossing points are essential to the existence of non-trivial solutions beyond Bloch waves and make contact with the topological properties of a graph $-$ In this case, a discretised quantum graph. They also make contact with the appearance of bound states around defects in solid state models. So, one may ask, how important is the role of topology for the existence of bound states in a system without loops? The results in this paper indicate that it can be actually very important, as such states appear as the consequence of a vertex. Including them is not optional \cite{17, 18} but mandatory.

Moreover, it is shown that nearest-neighbour chains in star configurations provide an exactly solvable model with analytic expressions for the energies and for the eigenvectors as functions of the sites in the chain. This can be most useful in our way to understand the origin of the binding effect.

Structure of the paper: In section 2, we present a numerical experiment in order to study the behaviour of bound states in the transition from a continuous system to quantum chains. Several examples are studied and the results are summarised in tables. Section 3 is devoted to exact solutions of the eigenvalue problem for star-shaped chains and a linear chain with a protuberance. Remarkably, the square of the energies for the latter case can be related to the golden ratio. In section 4, we give an experimental proposal using dielectric resonators in microwave cavities; such a system is simple enough to see how the effect can be easiliy observed for any type of waves. A discussion and some concluding remarks are given in section 5. 

\section{A numerical experiment}

In this section we show the results of a numerical experiment in which the continuous boundary value problem of crossing wires is discretised in order to solve the Schr\"odinger equation by the method of finite differences; such a discretisation is gradually taken to the limit of a very poor approximation in which the wires are one-cell thick. This is done with the purpose of studying the transition from quantum wires to quantum chains. We solve a few geometries of star-shaped wires such as Y, X and * crossings, and analyse the behaviour of bound states when the thickness of the wires is decreased to its minimum. The Laplace operator acting on a function $\nabla^2 f(x,y)$  is approximated by $f(x+1,y)+f(x,y+1)+f(x-1,y)+f(x,y-1)-4f(x,y)$. The matrix elements in terms of Kronecker deltas are used to diagonalise the Hamiltonian, with the possibility of obtaining all the eigenvectors and energies of the system. Different grid sizes and wire thickness were used in the calculations. The results are summarised in the tables \ref{table1}$-$\ref{table5}.

The results show essentially, that for the case of a fine discretisation mesh, a certain number of bound states (magic number) appears in the crossing of wires and that such a number can be related to the number of arms in the array. By keeping the width of the arms fixed in the process, we see that adding more arms implies an increase of space within the central region. As the process goes on, the system ressembles more and more a circular billiard and the corresponding bound states approach to the corresponding internal modes - Bessel functions. On the other hand, when the transition to a quantum chain takes place, the space in the central region reduces to a single point regardless of the number of arms. Yet, a bound state without nodes survives through the entire process. Moreover, another bound state with many oscillations appears in the opposite side of the energy band, due to the inherent discreteness of the system. There is, of course, a maximum frequency allowed. The following points should be noted:

\begin{itemize}

\item All the arrays we have used are finite; in strict terms one always has confined solutions. The identification of bound states comes from the variation of the total size of the system: The decay of densities along the arms and the corresponding energies are not significantly modified when the size is increased.

\item In the X junction, we have not reached the limit of $1$ cell thickness but we have shown the case of $3$ cells. This helps the visibility of the probability amplitudes in the central region. In all cases, we see that at least one bound state survives the process in the lower part of the spectrum. There is always a reflective symmetry of the spectrum due to discreteness, but this is only relevant for the case of true quantum chains. All the states in the upper part of the resulting band are highly oscillatory.

\item In the case of $40 \times 40$ arrays with six thick arms, we found two bound states below energy threshold $E_t$ and one above (but below the half-guide threshold, which is $4E_t$). The second state below threshold can be interpreted in two ways: As it is apparent from the figure, the state is antisymmetric with respect to only one axis. Therefore, due to the $C_6$ symmetry, we should have a six-fold degeneracy (the other states are obtained upon rotations). The fully antisymmetric state corresponding to this energy displays a number of peaks {\it inside\ } the interaction region, suggesting that a radial mode with zero density at the origin, can be accomodated in the central part. From pure geometrical arguments we see that this can be the case, as the side of an hexagon (wire thickness) coincides with its radius. A second way to understand the appearance of this state is to identify half of the hexagonal array with a slightly deformed T junction, which is known to accomodate one bound state. 

\item An observation in connection with symmetry is due. The chains share the same discrete rotational symmetry with the wires; we can find real antisymmetric solutions in a system with an even number of chains or arms. However, the question is, are the antisymmetric states in the chains bound as well? The answer is no: Lack of space.

\item In order to support our previous statements, all star-shaped arrays of linear chains with one central point shall be analysed with exact methods in section 3.

\end{itemize}

\newpage

\subsection{Numerical Results}

\begin{table}[h]
 
\caption{ \label{table1} The behaviour of bound states for three arms. First row: The numerical results for thick wires in a fine mesh. The energies of the states in the third column are given in units of the threshold value $E_t$. This value is defined by the first propagation mode of the straight guides. Second row: The effects of mesh coarsening. The energy Bloch band of an infinite, linear and discrete array is shown in the last column as two horizontal lines (red), while the energies of the system (blue curve) are given in decreasing order. The energies of the states in the third column are given in units of the center of the Bloch band $E_c$.}
\begin{tabular}{|m{3cm}| m{3cm} | m{5cm}| m{4.5cm}|} 

\hline Grid description &  Shape of the array & Bound states: Densities and energies & Comparison with Bloch band \\

\hline 
$40\times 40$ array. Wire thickness: $17$ sites. & \includegraphics[scale=0.25]{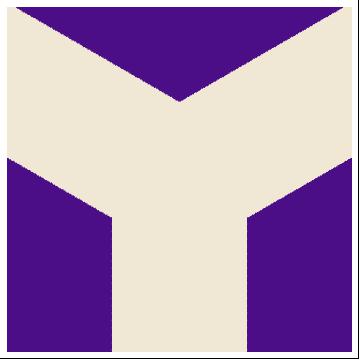} & \begin{tabular}{c} $E= 0.96 E_t$ \\ \includegraphics[scale=0.36]{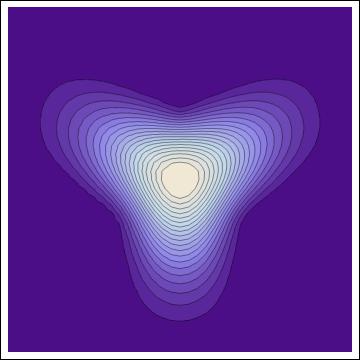} \end{tabular} & Not applicable. All bound states appear at low energies
\\
\hline 
$28\times 28$ array. Wire thickness: $1$ site. The third column shows a magnification of the central region. & 
\includegraphics[scale=0.25]{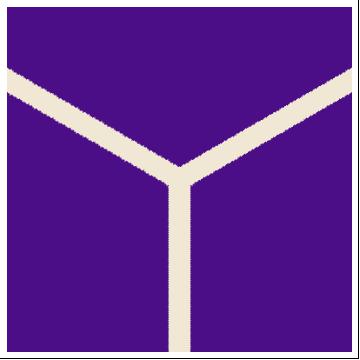} & \begin{center} \begin{tabular}{c} $E= 0.46 E_c$ \\ \includegraphics[scale=0.3]{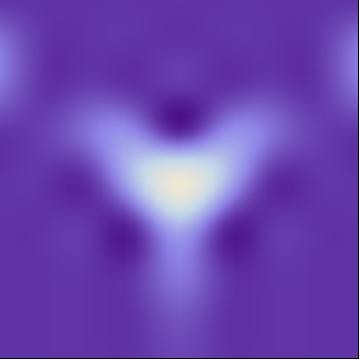} \\ $E=1.53 E_c$ \\ \includegraphics[scale=0.3]{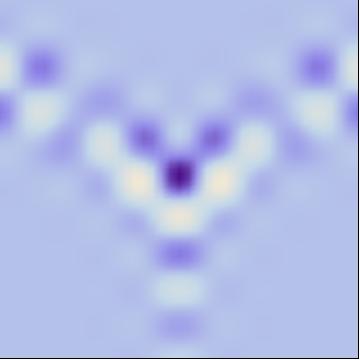} \end{tabular} \end{center} & \includegraphics[scale=0.6]{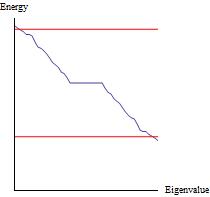} 
\\ 
\hline

\end{tabular}

\end{table}

\begin{table}

\caption{\label{table2} The behaviour of bound states for four arms. First row: Thick wires in a fine mesh. The energies of the states in the third column are given in units of the threshold value $E_t$. Second row: The effects of mesh coarsening. The energy Bloch band of an infinite, linear and discrete array is shown in the last column as two horizontal lines (red), while the energies of the system (blue curve) are given in decreasing order. The energies of the states in the third column are given in units of the center of the Bloch band $E_c$.}

\begin{tabular}{|m{3cm}| m{3cm}| m{5cm}|m{4.5cm}|} 

\hline Grid description &  Shape of the array & Bound states: Densities and energies & Comparison with Bloch band \\

\hline 
$40\times 40$ array. Wire thickness: $12$ sites. & \includegraphics[scale=0.25]{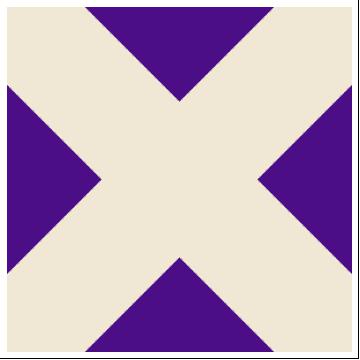} & \begin{center} \begin{tabular}{c} $E= 0.67 E_t$ \\ \includegraphics[scale=0.3]{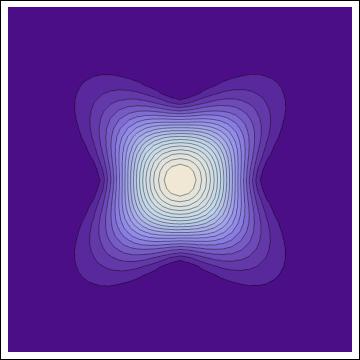} \\  $E= 3.73 E_t$ \\ \includegraphics[scale=0.3]{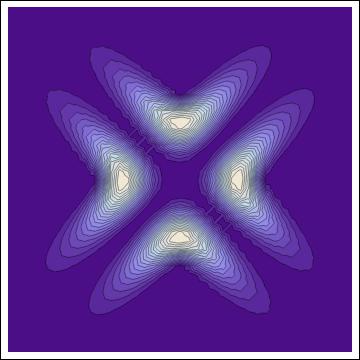} \end{tabular} \end{center} & Not applicable. All bound states appear at low energies
\\
\hline 
$28\times 28$ array. For better visibility, we have taken a wire thickness of $3$ sites. & 
\includegraphics[scale=0.25]{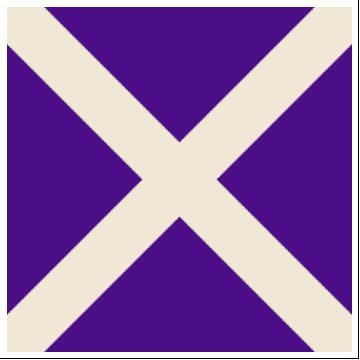} & \begin{center} \begin{tabular}{c} $E= 0.07 E_c$ \\ \includegraphics[scale=0.3]{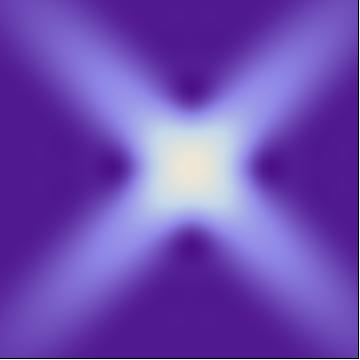} \\ $E= 1.92 E_c$ \\ \includegraphics[scale=0.3]{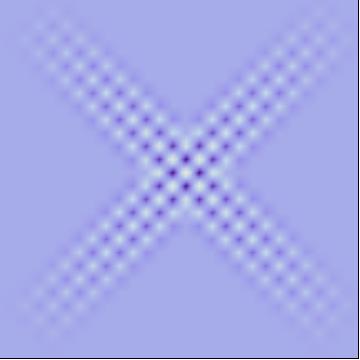} \end{tabular} \end{center} & \includegraphics[scale=0.6]{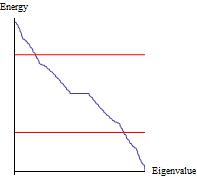} 
\\ 
\hline

\end{tabular}

\end{table}

\begin{table}

\caption{\label{table3} The behaviour of bound states for five arms. First row: The numerical results for thick wires in a fine mesh. The energies of the states in the third column are given in units of the threshold value $E_t$. This value is defined by the first propagation mode of the straight guides. Second row: The effects of mesh coarsening. The energy Bloch band of an infinite, linear and discrete array is shown in the last column as two horizontal lines (red), while the energies of the system (blue curve) are given in decreasing order. The energies of the states in the third column are given in units of the center of the Bloch band $E_c$.}

\begin{tabular}{|m{3cm}| m{3cm}| m{5cm}|m{4.5cm}|} 

\hline Grid description &  Shape of the array & Bound states: Densities and energies & Comparison with Bloch band \\

\hline 
$40\times 40$ array. Wire thickness: $9$ sites. & \includegraphics[scale=0.25]{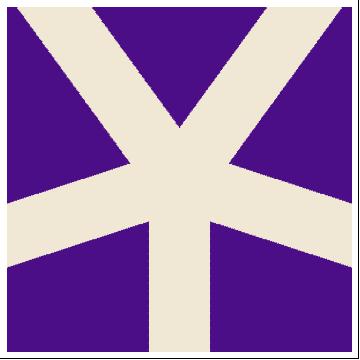} & \begin{tabular}{c} $E= 0.53 E_t$ \\ \includegraphics[scale=0.36]{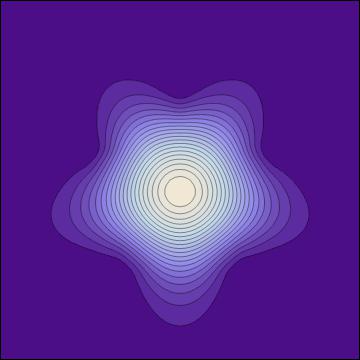} \end{tabular} & Not applicable. All bound states appear at low energies.
\\
\hline 
$28\times 28$ array. Wire thickness: $1$ site. The third column shows a magnification of the central region. & 
\includegraphics[scale=0.25]{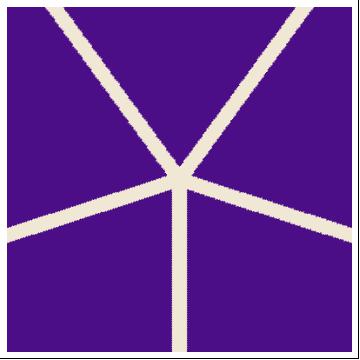} & \begin{center} \begin{tabular}{c} $E= 0.46 E_c$ \\ \includegraphics[scale=0.3]{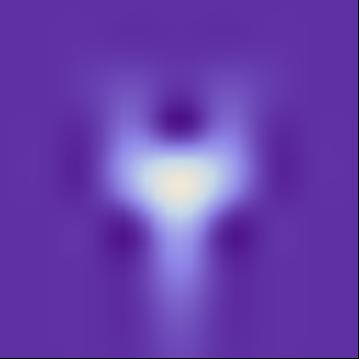} \\ $E= 1.54 E_c$ \\ \includegraphics[scale=0.3]{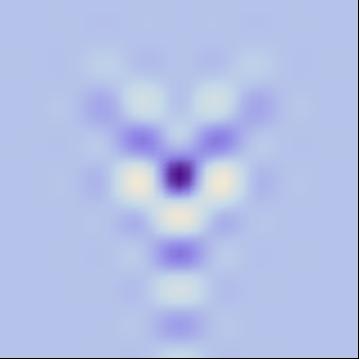} \end{tabular} \end{center} & \includegraphics[scale=0.6]{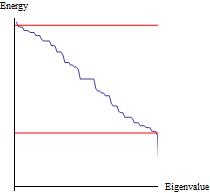} 
\\ 
\hline

\end{tabular}

\end{table}

\begin{table}

\caption{\label{table4} The behaviour of bound states for six arms. We show the numerical results for thick wires in a fine mesh, while the effects of mesh coarsening are reserved for the next table. The energies of the states in the third column are given in units of the threshold value $E_t$. This value is defined by the first propagation mode of the straight guides.}

\begin{tabular}{|m{3cm}| m{3cm}| m{5cm}|m{4.5cm}|} 

\hline Grid description &  Shape of the array & Bound states: Densities and energies & Remarks \\

\hline 
$40\times 40$ array. Wire thickness: $9$ sites. & \includegraphics[scale=0.25]{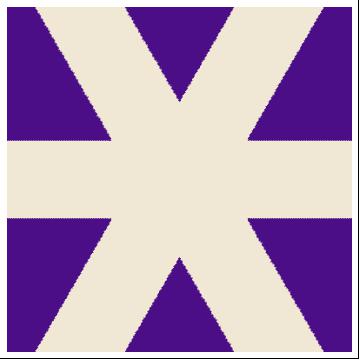} & \begin{tabular}{c} $E= 0.42 E_t$ \\ \includegraphics[scale=0.36]{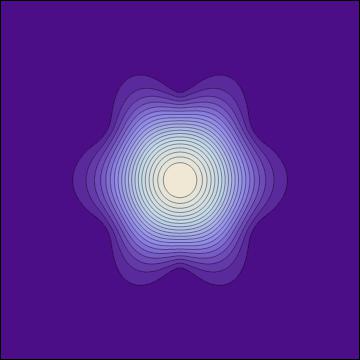} \\ $E= 0.91 E_t$ \\ \includegraphics[scale=0.36]{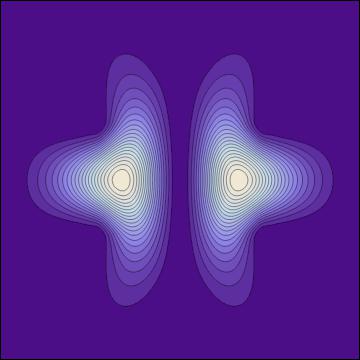} \\ $E= 3.37 E_t$ \\ \includegraphics[scale=0.36]{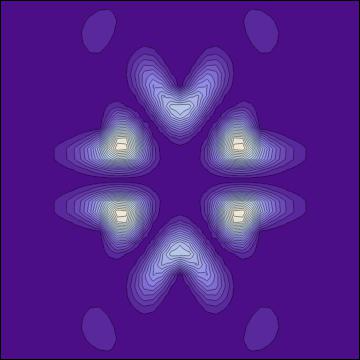} \end{tabular} &  All bound states appear at low energies. Two states are below threshold. The ground state is fully symmetric. The second state is antisymmetric and six-fold degenerate. The third state is fully antisymmetric and above threshold.
\\
\hline

\end{tabular}

\end{table}

\begin{table}

\caption{\label{table5} The behaviour of bound states for six arms. Here we give the effects of mesh coarsening. There are two bound states. The energy Bloch band of an infinite, linear and discrete array is shown in the last column as two horizontal lines (red), while the energies of the system (blue curve) are given in decreasing order. The energies of the states in the third column are given in units of the center of the Bloch band $E_c$.}

\begin{tabular}{|m{3cm}| m{3cm}| m{5cm}|m{4.5cm}|} 

\hline Grid description &  Shape of the array & Bound states: Densities and energies & Comparison with Bloch band \\

\hline
$28\times 28$ array. Wire thickness: $1$ site. The third column shows a magnification of the central region. & 
\includegraphics[scale=0.25]{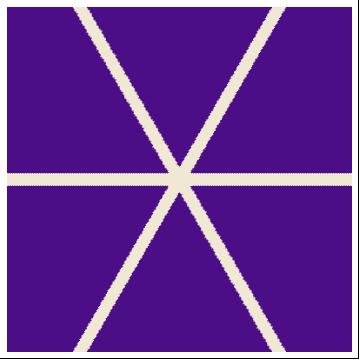} & \begin{center} \begin{tabular}{c} $E= 0.40 E_c$ \\ \includegraphics[scale=0.3]{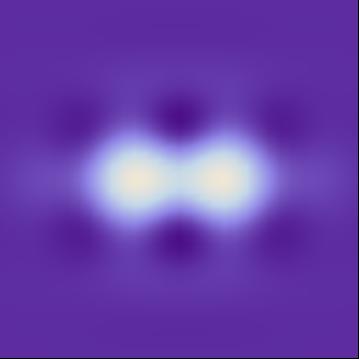} \\ $E= 1.59 E_c$ \\ \includegraphics[scale=0.3]{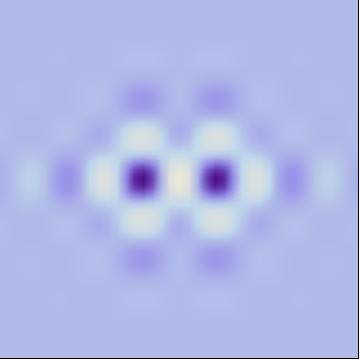} \end{tabular} \end{center} & \includegraphics[scale=0.6]{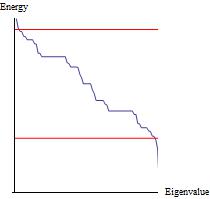} 
\\ 
\hline

\end{tabular}

\end{table}

\section{Bound states at a crossing point - Exact solutions}

As promised, here we show a method to extract analytical expressions for energies and eigenvectors of the bound states in a system of chains. For this purpose, we introduce a method of recursion in the site number of the corresponding arrays.

\subsection{Symmetric star-shaped arrays}

The hamiltonian of the tied chains or arms can be expressed as a matrix whose blocks are coupled to a single central point labeled $0$. There is one block for each arm. It is convenient to recognize that in the discretisation of the Laplacian we find an identity operator that can be removed from our hamiltonian for simplicity. The hamiltonian reads

\bea
\fl H= \left( \begin{array}{clll}  0 & \begin{array}{cccc} 1& 0& \cdots & 0 \end{array}  &  \begin{array}{cccc} 1& 0& \cdots & 0 \end{array} &    \cdots \\ 
 \begin{array}{c} 1\\ 0\\ \vdots \\ 0  \end{array} &  \left. \begin{array}{cccc} 0& 1& &  \\ 1& 0& 1 &  \\  & 1 & 0 &   \\  & &  & \ddots  \end{array} \right\}  \mbox{arm 1} & & \\

 \begin{array}{c} 1\\ 0\\ \vdots \\ 0 \end{array} & & \left. \begin{array}{cccc} 0& 1& &  \\ 1& 0& 1 &  \\  & 1 & 0 &   \\  & &  & \ddots  \end{array} \right\}  \mbox{arm 2} &  \\

 \vdots & & & \ddots 

\end{array} \right)_{\mbox{dim}=(pN+1)\times(pN+1)}
\label{1}
\eea
where the unspecified off-diagonal blocks are zero, $p$ denotes the number of arms and $N$ denotes the number of sites in each arm. The stationary Schr\"odinger equation $H \Phi=E\Phi$ is therefore a recurrence relation for the components of the wave vector $\Phi$, which we write as

\bea
 \Phi= \left( \begin{array}{c}  \phi_0 \\
 \left.\begin{array}{c} \phi_1\\ \vdots \\ \phi_N  \end{array} \right\}  \mbox{arm 1}  \\

 \left.\begin{array}{c} \phi_{N+1}\\ \vdots \\ \phi_{2N}  \end{array} \right\}  \mbox{arm 2}  \\

 \vdots

\end{array} \right)_{\mbox{dim}=pN+1}.
\label{2}
\eea
The recurrences are

\bea
\phi_{k+1}-E\phi_k + \phi_{k-1} = 0
\label{3}
\eea
along the arms, except for the components that are coupled to the centre $\phi_0$. We have the conditions

\bea
\begin{array}{llll}
2 \leq k \leq N  &\mbox{and}&  \phi_{2}-E\phi_1 + \phi_{0} = 0  &\mbox{for arm $1$} \\ 
N+2 \leq k \leq 2N &\mbox{and}&  \phi_{N+2}-E\phi_{N+1} + \phi_{0} = 0 &\mbox{for arm $2$} \\
2N+2 \leq k \leq 3N &\mbox{and}& \phi_{2N+2}-E\phi_{2N+1} + \phi_{0} = 0  &\mbox{for arm $3$} \\
\vdots && \vdots &  \end{array}
\label{4}
\eea
The coupling elements appearing in the $0$-th row produce the relation

\bea
\phi_1 + \phi_{N+1} + \phi_{2N+1} + \cdots = E \phi_0,
\label{5}
\eea
and this, in turn, can be solved by considering eigenvectors $\Phi$ corresponding to real symmetric representations of the group $C_{pV}$, \ie the symmetry group of the problem  (the antisymmetric solutions can be studied too, but they do not yield interesting results for our purposes). This implies that (\ref{5}) is solved by

\bea
\phi_1 = \phi_{N+1} = \phi_{2N+1} = \cdots = \frac{E \phi_0}{p} 
\label{6}
\eea
fixing therefore the first component for each arm. Now we turn to the solutions of (\ref{3}). Since the arms are coupled to the centre, but not directly between themselves, we can start by solving the recurrence for only one arm (say, arm $1$) and repeat the solution for the rest of the system, with the proviso that (\ref{6}) contains already the (indirect) interaction between chains. We consider

\bea
\phi_{k+1}-E\phi_k + \phi_{k-1} = 0, \qquad 1 \leq k \leq N
\label{7}
\eea
for the first arm, and solve it by writing

\bea
\phi_k = A_+  (\mu_+)^k + A_-  (\mu_-)^k
\label{8}
\eea
with $A_{\pm}$ a pair of coefficients to be determined, and $\mu_{\pm}$ given by the two roots of the equation $\mu^2 - E \mu + 1=0$, \ie

\bea
\mu_{\pm} = \ahalf \left( E \pm \sqrt{E^2-4} \right)
\label{9}
\eea
Usually, these relations do not provide directly the energies (e.g. in periodic chains), but they cast the coefficients $A_{\pm}$ in terms of the eigenvalues $E$ according to the choice $A_{+}=0$ or $A_{-}=0$. In our case, however, we can exploit the fact that $\phi_1$ has been fixed already by the interaction in the form (\ref{6}), leading to

\bea
\phi_1 =  A_+  \mu_+ + A_-  \mu_- = \frac{E \phi_0}{p} = \frac{E}{p} \left(  A_+  + A_-  \right)
\label{10}
\eea
and this relates the energy $E$ with the number of arms $p$ and the coefficients $A_{\pm}$. But before solving for $E$, let us analyse the general properties of the relations obtained so far. From (\ref{9}) we find that if $E$ lives in the band $[-2,2]$, then $\mu_{\pm}$ are complex conjugate to each other and live in the unit circle, leading to an oscillatory behaviour of $\phi_k$ as a function of $k$. On the other hand, if $|E|>2$, the roots $\mu_{\pm}$ become real and satisfy $| \mu_+|>1$, $| \mu_-|<1$ for $E>0$ and $| \mu_-|>1$, $| \mu_+|<1$ for $E<0$. The general form of $\phi_k$ in (\ref{8}) shows that, for infinite chains, square summable functions are possible whenever $\mu_+$ or $\mu_-$ appear alone in the expression, as the other one grows exponentially with $k$. Therefore $A_+=0$ and $A_-=\phi_0$ or $A_-=0$ and $A_+=\phi_0$. In both cases we have that (\ref{10}) gives the energies in terms of $p$ alone, as $\phi_0$ cancels out:

\bea
\phi_0 \mu_{\pm}= \frac{E \phi_0}{p}
\label{11}
\eea
or
\bea
 \ahalf \left( E \pm \sqrt{E^2-4} \right)= \frac{E}{p}
\label{12}
\eea
The two solutions are finally

\bea
E_{\pm} = \pm \frac{p}{\sqrt{p-1}}, \qquad p\geq 2.
\label{13}
\eea
This is indeed a very simple expression for the energies of the two bound states lying outside of the band $[-2,2]$. For $p=2$, we recover the linear infinite chain and $E=\pm 2$, corresponding to $|\phi_k|= \mbox{constant}$ (the state is not bound). For $p>2$, the eigenvectors of the two bound solutions are

\bea
\phi^{\pm}_k = (\pm)^k (p-1)^{-k/2}  \phi_0
\label{14}
\eea
which decay exponentially fast. The constant $\phi_0$ is obviously fixed by normalization. The positive energy state oscillates with maximal frequency as the sign of $(\mu_-)^k$ flips from site to site  (here we should recall that the removal of the identity in the discrete Laplacian shifts the spectrum downwards, allowing therefore negative energies as well).

\begin{figure} \begin{center} \begin{tabular}{cc} \includegraphics[scale=0.8]{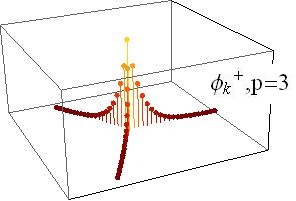} &  \includegraphics[scale=0.7]{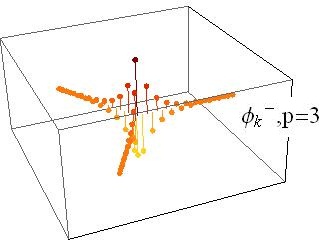} \\
 \includegraphics[scale=0.8]{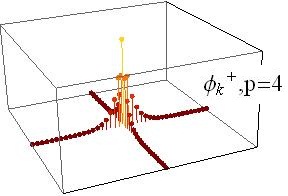} &  \includegraphics[scale=0.7]{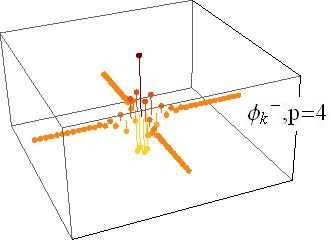} \\
 \includegraphics[scale=0.8]{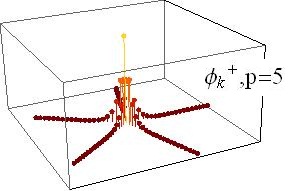} &  \includegraphics[scale=0.7]{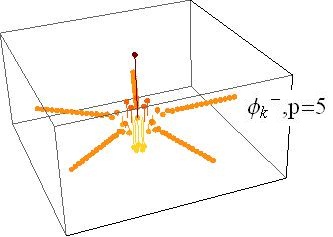}
\end{tabular} \end{center} 

\caption{ \label{fig:4} Bound state vectors as a function of site number. In the left column we show the nodeless solutions for arrays with three arms (first row), four arms (middle row) and five arms (third row). In the right column we show the higly oscillatory solutions for the corresponding value of $p$. Each point is coloured according to the value of $\phi_k$, for a better visibility.} \end{figure}

\begin{figure} \begin{center} \begin{tabular}{c} \includegraphics[scale=0.7]{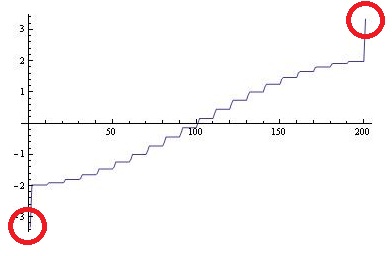}
\end{tabular} \end{center} 

\caption{ \label{fig:5} Numerical energy spectrum for an array of ten arms, each arm having twenty sites. Here we see that the two energies of the bound states are well separated from the rest of the band: $E=\mp 3.33$ for the levels labeled by $1$ and $200$ respectively. The energies within the band form plateaux due to an approximate ten-fold degeneracy. Despite of the finiteness of the chains ($N=20$), the results come out as predicted by the theory.} \end{figure}

We include some plots (see figures \ref{fig:4} and \ref{fig:5}) showing the eigenstates and spectrum for a few examples.

\subsection{A linear chain with a stem or protuberance}

In the right panel of figure \ref{fig:2}, we have shown a caricature of a waveguide with a bulge as a linear chain with an additional site stemming out of it. In order to complete our study, it is convenient to treat this configuration as well. We proceed to give the exact solution of the corresponding tight-binding problem using the tools we have developed above. Interestingly, the bound state energies can be given in terms of the golden ratio.

From the previous treatment and from figure \ref{fig:2}, we can see that this system corresponds to $p=3$, with two infinite arms attached to the centre and one finite  arm with $N=1$. The configuration is symmetric only with respect to parity (dihedral group) around the centre, so we start from (\ref{5}) and one step before the fully symmetric choice (\ref{6}) is made. Instead of (\ref{6}), we have the condition $\phi_1=\phi_{N+1}$ for the two symmetric infinite arms. The energy equation for the one-site arm (the component $2N+1$) and the equation for the $0$th row of $H$ read

\bea
\phi_0=E\phi_{2N+1} \qquad \mbox{and} \qquad \phi_{1}+ \phi_{N+1} +\phi_{2N+1}=E\phi_0
\label{inter1}
\eea
respectively. Setting $\phi_{N+1}=\phi_{1}$ and eliminating $\phi_{2N+1}$ in (\ref{inter1}) gives the condition

\bea
\phi_1 = \left( \frac{E^2-1}{2E} \right) \phi_0.
\label{inter2}
\eea
Now we repeat the process in the previous section, by solving the recurrence for one infinite arm (again arm $1$). We must choose only one of the roots $\mu_{\pm}$ in the solution $\phi_k$, as the latter must be square summable. The resulting $\phi_0, \phi_1$ can be substituted back in (\ref{inter2}), finding an expression in terms of the energies alone

\bea
1\pm\sqrt{E^2-4}E = 0
\label{inter3}
\eea
where the upper and lower sign correspond to the two possible choices $\mu_{\mp}$. The solutions are given thus in terms of the {\it golden ratio\ }:

\bea
E=\pm \sqrt{2\tau + 1}
\label{inter4}
\eea
where $\tau=(1+\sqrt{5})/2$. Substituting back in $\mu_{\pm}$ given by (\ref{9}), the two evanescent solutions become

\bea
\phi^{\pm}_k =  (\pm)^k \left( \frac{\sqrt{2\tau+1}-\sqrt{2\tau-3}}{2}\right)^k \phi_0
\label{inter5}
\eea
We can verify that the two bound state energies lie outside of the band $[-2,2]$ as before, but they are rather close to the edges. As to the eigenvectors, we have a behaviour similar to the completely symmetric case: One nodeless function with exponential decay and one oscillatory solution with its corresponding exponential envelope. In this case, the numerical value $|\mu|=0.786151...$ results larger than $|\mu|=1/\sqrt{2}$ for the three symmetric infinite arms, which implies that (\ref{inter5}) decays slower than any of the previous cases with $p \geq 3$. Therefore, the binding capabilities of a discrete one-dimensional protuberance are evident, although they are very weak.

\section{Experimental realisation}

In this section we indicate the realisation of our system in tight-binding arrays using microwave cavities. In recent years, the works by Mortessagne and collaborators have shown that arrays of dielectric resonators can be used to demonstrate a number of effects found in two-dimensional solid state physics \cite{19}, by considering ordered and integrable systems \cite{20} as well as disordered systems \cite{21}. These effects range from the presence of band gaps to Dirac points in hexagonal arrays \cite{19}, to Anderson localization in disordered configurations \cite{21}. The idea is very simple: We work with an effective two-dimensional description of a scalar wave given by a component of the electric field in a microwave cavity of constant height and made of a good conductor, as prescribed by St\"ockmann \cite{22}. In such a cavity, dielectric resonators of cylindrical shape (made of a ceramic material with permitivity $\epsilon \sim 37$) can be placed and arranged in a most flexible way. Each of the resonators emulates a site of the tight-binding chains described before. Here, of course, only one isolated resonance of a cylinder enters into play and it is considered to be the same for all cylinders as they are approximately identical. Then, the distance between cylinders is fixed such that the coupling between field resonances extends only to nearest neighbours. This has been carefully demonstrated experimentally \cite{19} by varying the distance between a pair of interacting cylinders of approximately $8$mm in diameter and with a narrow resonance around $\omega_0=6.66$GHz. There, it was verified that the level splitting of such a two-level system decreases exponentially with the distance of separation. Such a splitting is given directly by the off-diagonal coupling of the system $\Delta$. For example, $\Delta$ can be measured to be $70$MHz, for a separation of $3$mm between resonators.
 
 \begin{figure} \begin{center} \begin{tabular}{c} \includegraphics[scale=0.7]{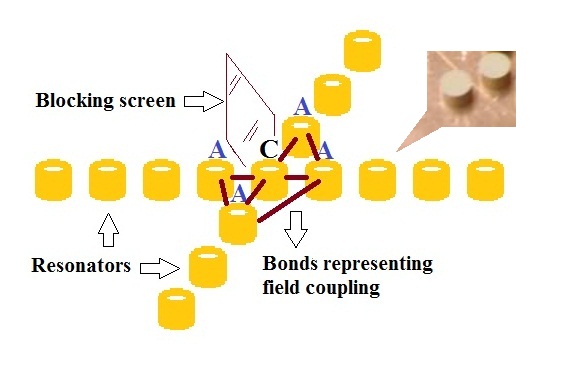}
\end{tabular} \end{center} 

\caption{ \label{fig:6} An array of dielectric resonators in the form of cylinders, each of them supporting an isolated and very narrow resonance. The separation distance between them can be adjusted such that only the nearest-neighbour coupling becomes important. The bonds shown in red represent some of those couplings for the resonators in the central region: The electromagnetic field in the cylinders A is supposed to feel the field in C, as well as the field of their counterparts in adjacent arms. This would spoil the realisation of our model unless a blocking screen made of a conducting material is placed between the arms. The extra couplings in the center do not preclude the existence of bound states, but the corresponding model becomes more involved.} \end{figure}

One technical aspect that should be covered in such a realisation, is the possibility of a non-vanishing coupling between second neighbours at the crossing point of the chains. See figure \ref{fig:6}. The field in the central cylinder C is supposed to see only the resonances from cylinders A. This is indeed the case if the distance between them is not too small. However, the field in a cylinder A might couple to its counterpart in the adjacent arm. In order to block such an interaction, one may simply put a screen of a conducting material. For example, a small copper plate.

We should stress that the absence of a blocking screen would modify the interaction only in the central region by allowing extra bonds. Therefore, we would expect again the existence of bound states in such an uncrontrolled situation. Nevertheless, the number of bound states and their energies can no longer be predicted with our model.

We expect the signature of the states in question in the part of the spectrum lying outside of the Bloch band corresponding to linear periodic chains. The width of the band is determined by twice the coupling $\Delta$ between nearest neighbours, a typical value being $140$MHz as mentioned before. The transmission between antennas can be measured as indicated in \cite{19}, both as a function of the injected frequency and the distance between them. We expect that for a pair of antennas placed close to the cylinders in the central region of the array, there should be two peaks in the transmission well separated from the resonance of an individual, isolated cylinder. According to our theory, they should appear at frequencies $\omega_0 \pm p\Delta/\sqrt{p-1}$. As the antennas are moved along the arms and their distance is increased, the coupling between them should vanish exponentially, irrespective of the relative position between antennas and resonators. This shall be the clear indication of a bound state.

\section{Summary and Discussion}

Let us summarise our results. We have shown that unusual bound states appearing in star-shaped junctions may diminish in number under a discretisation and dimensional reduction of the system, \ie quantum chains. Only one bound state remains in the lower part of the spectrum, but another one appears due to the discreteness of the system. Therefore, the magic numbers of the array depend strongly on the conditions given above. The remaining bound states depend crucially on the existence of a vertex. The exact solutions of the problem exhibit the binding mechanism: The exponential decay of the eigenfunctions is due to values of the energy outside of the Bloch band (and their effect on $\mu_{\pm}$). We have proposed an experiment in order to visualise the effect in simple terms. Arrays of resonators can be incarnated in many forms, the dielectric cylinders supporting microwave resonances being only one example. 

At this point we can discuss some consequences of our findings. Let us focus first on the continuous case. In symmetrical star-shaped wires, the thickness of the leads acts as a natural scale parameter, such that a variation of the distance between the walls of the guides can be reinterpreted as a scale transformation of the variables $x,y$. This only holds for a special type of geometries. The structure of the spectrum is therefore invariant under scaling, in the sense that the number of bound states remains intact. Let $x=\lambda x', y=\lambda y'$. The Helmholtz equation

\bea
\left[\nabla^2_{x,y} + E \right] \phi(x,y)=0
\label{eq1}
\eea
transforms trivially as

\bea
\left[\nabla^2_{x',y'} + E\lambda^2 \right] \phi(\lambda x',\lambda y')=0,
\label{eq2}
\eea
and the energy simply scales with $\lambda^2$, which is related to the thickness. Interestingly, in some applications of one-dimensional quantum mechanics and quantum field theories such as quantum graphs \cite{17}, bound states in the absence of loops (and in the absence of additional potentials) are put by hand or disregarded at all. Of course, this type of systems are {\it born\ } as one-dimensional and with zero threshold, whereas in guides with a certain thickness we always have a fixed number of bound states during the limiting process $\lambda\rightarrow 0$. The dimensional reduction is thus a version of the effects described many years ago by da Costa \cite{23}, where dimensionality combined with curvature gave rise to purely quantum potentials. 

Now we turn to the discrete case. In contrast with the previous observations, our examples of quantum chains can be considered to be a one-dimensional idealization of a physical system {\it ab initio\ }and where the bound states are unavoidable. The low dimensionality and the non-trivial topology are accompanied by discreteness, the latter being an important ingredient. This is a clear indication that a theory of quantum graphs in lattices should incorporate the results of the present paper.

\ack

The author would like to express his gratitude to CONACyT {\it programa de repatriaci\'on, convocatoria 2011,\ }for financial support.

\end{document}